# Time-domain feature extraction for target-specificity in Photoacoustic Remote Sensing Microscopy

Nicholas Pellegrino,[1,2] Benjamin R. Ecclestone,[1,3] Paul Fieguth,[2] Parsin Haji Reza[1,*]

[1]PhotoMedicine Labs, University of Waterloo, 200 University Ave W, Waterloo, ON N2L 3G1
[2]Vision and Image Processing Lab, University of Waterloo, 200 University Ave W, Waterloo, ON N2L 3G1
[3]illumiSonics Inc., 22 King Street South, Suite 300, Waterloo, ON, N2J 1N8
*Corresponding author: phajireza@uwaterloo.ca

**Photoacoustic Remote Sensing (PARS) microscopy is an emerging label-free optical absorption imaging modality. PARS operates by capturing nanosecond-scale optical perturbations generated by photoacoustic pressures. These time-domain (TD) modulations are usually projected by amplitude to determine absorption magnitude. However, significant information on the target's material properties is contained within the TD signals. This work proposes a novel clustering method to learn TD features which relate to underlying biomolecule characteristics. This technique identifies features related to constituent biomolecules, enabling single-acquisition virtual tissue labelling. Colorized visualizations of tissue are produced, highlighting specific tissue components. This is demonstrated on freshly resected murine brain tissue, clearly discerning structures including myelinated and unmyelinated neurons (white and gray matter) and nuclear structures.**

In recent years, interest in label-free microscopy techniques has risen dramatically, driven by the desire to observe cells and tissues in their native environments. Labeling fundamentally alters tissue chemistry, introducing undesired experimental complexity and confounding factors. Hence, label-free approaches to visualize the intrinsic contrast of cells and tissues without extensive modification are viewed as an avenue to revolutionize biological understandings.

One method, Photoacoustic Remote Sensing (PARS) microscopy [1] has emerged as a powerful label-free technique. PARS employs an all-optical architecture to visualize endogenous absorption contrast [1-5]. In practice, many biomolecules exhibit unique absorption spectra, providing means to selectively image individual chromophores [2,] directly within tissue specimens. Subsequently, PARS may provide label-free images of biomolecules such as hemoglobin [1], DNA [2-4], and lipids [2], in unprocessed surgical resections [2,3], and living animal subjects [1]. In PARS, a pulsed excitation laser is used to deposit focused optical energy into a specimen. Absorption of this energy results in localized heating and, the generation of photoacoustic pressures [1-5]. The localized thermal and pressure perturbations generate nanosecond-scale variations in the specimens' local optical properties [1-5], which are captured as backscattering intensity fluctuations in a co-focused PARS detection laser [1-5]. Hence, each excitation event results in a PARS time-domain (TD) signal, containing the nanosecond-scale absorption induced pressure and heat modulations signals.

In conventional PARS implementations, a single characteristic intensity value is extracted from each TD signal to visualize the total absorption magnitude at each point [2]. For example, TD amplitude, computed as the difference between the maximum and minimum of the TD signal, is commonly used to represent the absorption magnitude [1,2]. However, PARS signals contain significant time-resolved optical modulation information, which potentially carries additional specimen information, beyond the traditional absorption magnitude [5,6]. Indeed, the underlying physics of the PARS mechanism suggest that the time-evolution of PARS signals may be dictated by material properties such as the density, heat capacity, and acoustic impedance [5,6]. However, directly extracting such information from the TD signals is not necessarily straightforward. Hence, previous approaches aimed to assess attributes of the PARS TD signals independent of their source [7,8]. For example, Kedarisetti et al. proposed a method to leverage PARS TD frequency content as a means of inferring characteristics of the imaging target [7].

This work takes a different approach, proposing an unsupervised (clustering) method to learn time-domain features related to the underlying specimen. The proposition is that material-specific (biologically significant) information may be inferred from the time-domain signal content. For a given biomolecule with constant material properties, the PARS TD signals may have specific shapes. However, signals from a given target may vary in amplitude (e.g. based on concentration) and may suffer from noise. Hence, a method is required to determine constituent time-domain features that capture the material-specific information of the underlying tissue target, regardless of the noise and amplitude variation present in the TD signals. This is accomplished by clustering signals by shape, thus learning an associated prototype for each cluster.

Here, an optimized clustering method proposed by Pellegrino et al. is utilized [9]. Briefly, the method is based on K-Means clustering, where the main modification is an alternative method for computing cluster centroids. Measured signals are treated as vectors, where the vector angle is analogous to signal shape. The distance or difference between TD signals is the sine of the subtended angle, such that orthogonal signals have maximal distance and scaled or inverted signals have zero distance. Cluster centroids are then calculated as the first principal component of the union set of each cluster and its negative, causing the learned centroids to be robust to noise. Once the TD features (centroids) are learned, corresponding feature amplitudes are extracted by performing a change-of-basis from the time- to feature-domain.

This approach is applied directly to PARS images of unprocessed resected murine brain specimens. Here, a broadly absorbed UV excitation (266 nm) targets several biomolecules such as collagen, elastin, myelin, DNA, and RNA with a single excitation. Subsequently, the clustering approach is used to create enhanced absorption contrast visualizations and to extract biomolecule-specific features from the TD signals.

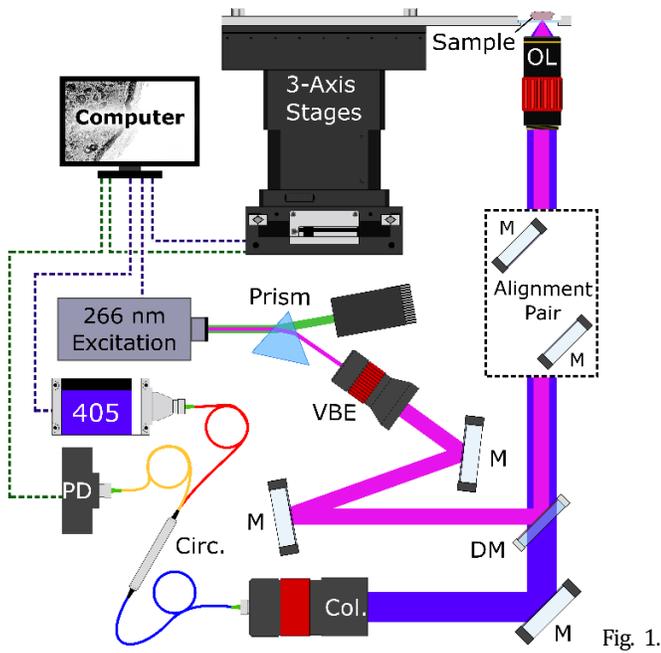

Fig. 1. Experimental setup. Component labels are: dichroic mirror (DM), variable beam expander (VBE), quarter waveplate (QWP), polarized beam splitter (PBS), long-pass filter (LP), collimator (C).

UV excitation is provided by a 50 kHz 266 nm laser (WEDGE XF 266, Bright Solutions) (Figure 1). The excitation is spectrally filtered with a prism, then expanded before combination with the detection beam. Detection light is provided by a continuous-wave 405 nm OBIS LS laser. The detection is fiber-coupled through the circulator, collimated, then combined with the excitation beam via a dichroic mirror. Combined excitation and detection are co-focused through a UV-transparent window onto the specimen. Back-reflected light from the sample returns to the collimator and circulator by the same path as forward propagation. The circulator re-directs backscattered light to the photodiode capturing the nanosecond-scale intensity modulations. During image acquisition, the stages raster scan the specimen over the objective lens, while the excitation pulses continuously. Analog photodiode output is captured for each excitation event using a high-speed digitizer, forming the PARS TD signals. Based on the stage position, each PARS TD signal is then mapped to a pixel in the final image.

Although conventionally PARS pixel values are defined by a single amplitude, the goal of this work is to enhance interpretability by extracting unique and more meaningful features based on the proposed K-means method. Different strategies to leverage the resulting features are explored, depending on number of extracted clusters.

If only a single feature is requested ($K = 1$), the clustering algorithm yields a feature, containing the TD shape similar to all tissue components. This feature can then be used as the basis for matched filtering, a technique designed to optimally extract the amplitude of known signal shapes with additive noise [8,10]. This feature provides a robust noise-resistant method for determining absorption amplitude or pixel "brightness". Applied in tissues (Figure 3(b)), this extraction provides a very substantial improvement in structural image quality and noise suppression compared to traditional TD amplitude projection (Figure 3 (a)).

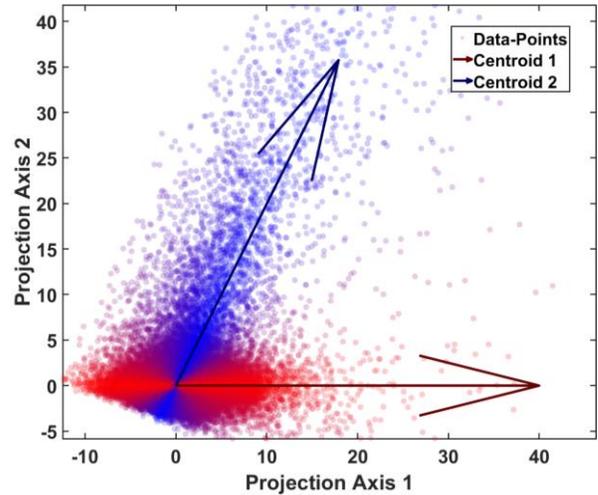

Fig. 2. Data-points (TD signals) are projected onto a plane defined by the two learned centroids. Points are colored according to the proportional signal content from each learned feature. The two arrows represent the learned centroids (i.e., characteristic time-domain signal shapes).

If additional clusters are requested ($K > 1$), tissue-specific features begin to emerge. In this case, the feature amplitudes at each pixel are extracted by performing a change-of-basis from the time-domain to the feature-domain. To visually illustrate the efficacy in learning features, a set of time-domain signals was clustered for $K = 2$ requested features. Figure 2 plots the projection of the high-dimensional time-domain data onto the plane containing the learned features, from which it is possible to visualize the connection between the underlying TD signals (dots) relative to the inferred features (arrows). In the visualization, each point is colored proportionally to the signal content attributed to the constituent features.

Further visualizations (Figure 3 (c)) are generated for resected murine brain tissues using three features ($K = 3$). The extracted feature amplitudes are mapped to the independent red, green and blue (R,G,B) color channels to form a colorized visualization. Hence, the pixel color represents the proportional mixture of each of the three features' contribution to the time domain signal, while the intensity represents the total magnitude of absorbed energy.

The $K = 3$ colorization (Figure 3 (c)) demonstrates the potential of the proposed technique in recovering biomolecule-specific information. Structures of singular myelinated neurons (white matter) from the brain stem are illustrated in pink, projecting into the brain. Concurrently, unmyelinated neurons (gray matter) appear on the right side of the frame in green. Finally, nuclear structures scattered throughout the brain tissues appear in white. The colorization of the different brain tissue constituents is explored further in Figure 4. Here, three different regions of the brain tissues (white matter (Figure 4 (a)), gray matter (Figure 4(c)) and the transition or boundary between white and gray matter (Figure 4(b))) were selected based on macroscopic inspection. Each unique region was imaged with the PARS microscope, before being colorized using the same $K = 3$ model as was used in Figure 3(c). In each of the selected regions, the TD colorization highlights identical biomolecule-specific structures as those identified in the initial colorized image (Figure 3(c)).

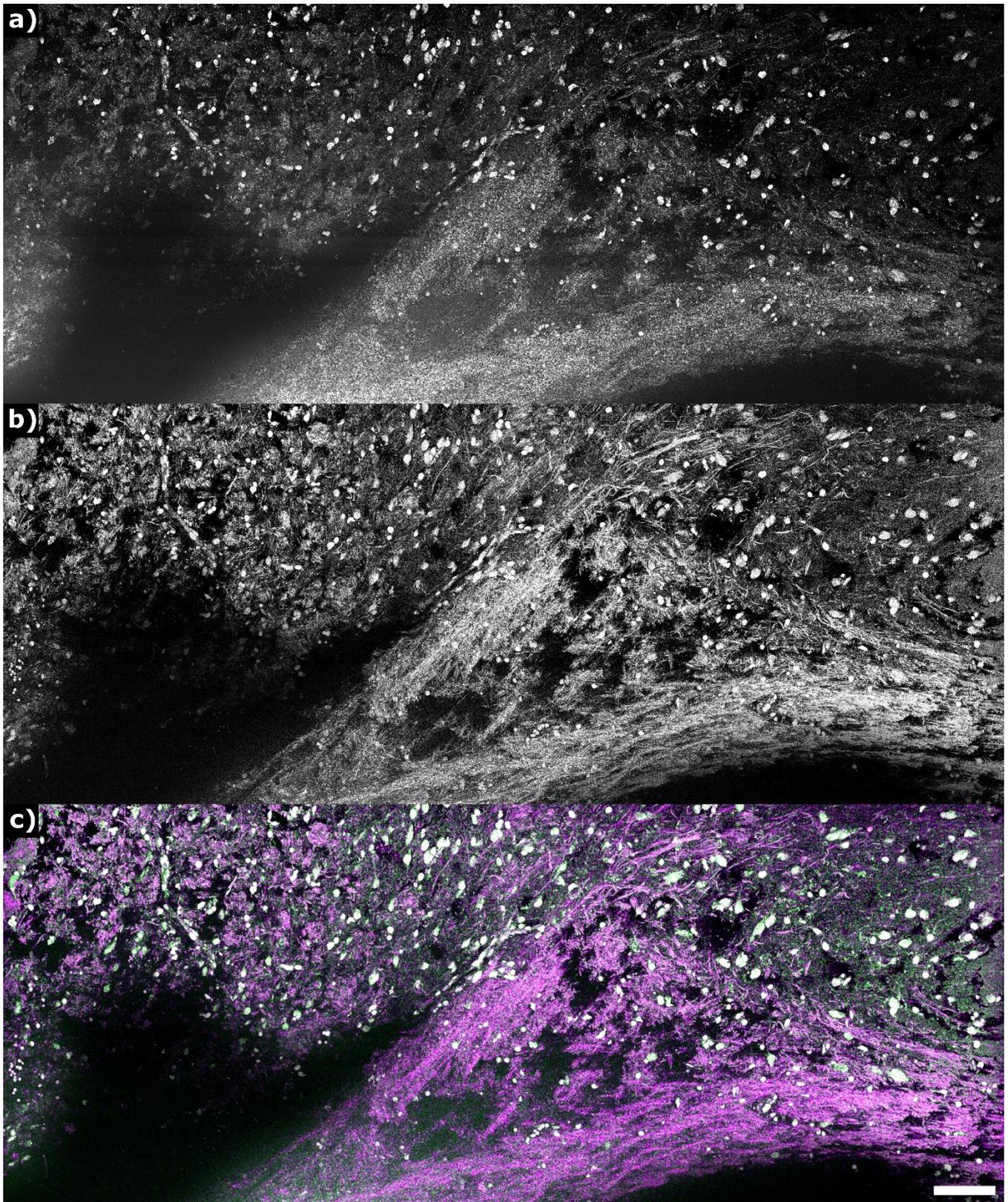

Fig. 3. PARS imagery of freshly resected murine brain tissue processed using three schemes. Panel a) shows the standard time-domain projection. Panel b) shows the matched filter projection based on a single learned feature ($K = 1$). Structural clarity and noise suppression is strongly improved here relative to panel a). Finally, panel c) combines extracted feature amplitudes for three ($K = 3$) learned features. The features are combined to produce a color image: Feature 1 maps to red, Feature 2 maps to green, Feature 3 maps to blue. Quite remarkable structures and tissue differentiation are clearly visible in this image, relative to that of panels a) and b). Scale Bar: 100 $\mu m$.

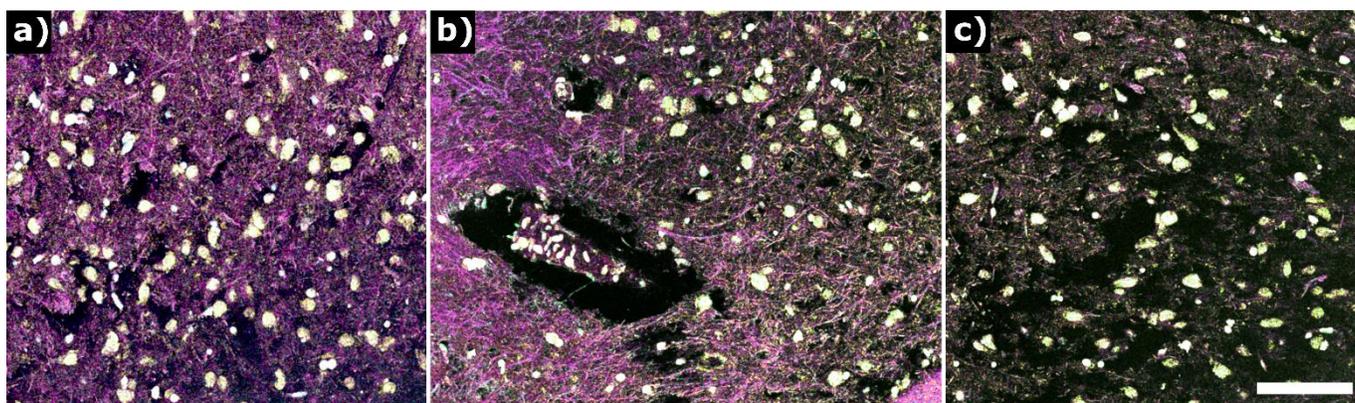

Fig. 4. Different sections of tissues visible with the PARS imagery of freshly resected murine brain tissue. a) Densely interwoven myelinated neurons with interspersed nuclei of the white matter region. b) Transition between white and gray matter where myelinated neurons penetrate into the gray matter. c) Gray matter region of the brain featuring predominately unmyelinated neurons with scattering nuclei. Scale Bar: 100 $\mu m$.

The implications of these results are twofold. First, the PARS TD signals contain sufficient information to identify biomolecules based on their clustered TD features. Second, such characteristics are transferrable across images of different tissue specimens. Feature identification was performed on an initial specimen, then transferred to others, producing similarly convincing results. Moreover, this technique offers unique advantages since the clustering approach requires no prior information, with the exception of the number of clusters, $K$. Although results were shown for $K = 1$ and $K = 3$, there is no upper limit to possible choices of $K$, except that visualization and interpretation do become increasingly challenging for $K > 3$. The learning is unsupervised, meaning that the method is applied blindly across the signals captured within the specimen of interest and does not require a-priori labeled or supervised data, especially beneficial in complex specimens such as the resected brain tissues explored here. The challenge is that blindly clustering for a pre-selected number of features does not guarantee that a singular biomolecule/tissue type will be isolated per feature. Each cluster simply targets a unique characteristic of the PARS TD signals, which may be used to highlight distinct tissue components.

In practice, this work shows that biomolecules may be visualized based on their PARS TD characteristics. This method may enable a single broadly absorbed excitation source to provide otherwise inaccessible material specificity, while simultaneously targeting the optical absorption of several biomolecules. This could enable enhanced absorption contrast visualizations, acquired in a fraction of the time compared to analogous multiwavelength approaches [2,4]. This enables several new avenues for label-free PARS microscopy by adding additional dimensions to the absorption contrast, vastly expanding the potential for biomolecule specificity.

**Funding.** Natural Sciences and Engineering Research Council of Canada (DGECR-2019-00143, RGPIN2019-06134); Canada Foundation for Innovation (JELF #38000); Mitacs Accelerate (IT13594); University of Waterloo Startup funds; Centre for Bioengineering and Biotechnology (CBB Seed fund); illumiSonics Inc (SRA #083181); New frontiers in research fund – exploration (NFRFE-2019-01012).

**Acknowledgments.** Authors NP and BRE contributed equally to the composition of this work. The authors thank Jean Flanagan, and the staff of the University of Waterloo Central Animal Facility for their help in procuring and preparing animal tissue specimens for this study.

**Disclosures.** BRE: illumiSonics Inc. (I,E,P), PHR: illumiSonics Inc. (F,I,E,P). NP and PF declare no conflicts of interest.

**Data availability.** Data underlying the results presented in this paper are not publicly available at this time but may be obtained from the authors upon reasonable request.